\documentclass[12pt]{article}
\usepackage{latexsym,amssymb,amsmath}
\numberwithin{equation}{section}

\begin{document}

\title
{On a class of gauge theories}
\author
{E.K. Loginov\footnote{{\it E-mail address:} ek.loginov@mail.ru}
\medskip\\
\it Department of Physics, Ivanovo State University\\
\it Ermaka Street 39, Ivanovo 153025, Russia}
\date{}
\maketitle

\begin{abstract}
We give a framework to describe gauge theory in which a nonassociative Moufang loop takes the
place of the structure group. The structure of such gauge theory has many formal similarities
with that of Yang-Mills theory. We extend the gauge invariance to this theory and construct an
on-shell version of $N=1$ supersymmetric gauge theory.
\end{abstract}

\section{Introduction}

In the past few ten years there have been many attempts to incorporate the unique algebra of
octonions into physics. From the early 1970s and up to the present time, octonions have been
applied with some success to different important problems such as quark confinement and grand
unified model, see Ref.~1. Starting from the 1980s, new applications of octonions in physics
were found, the instanton problem, supersymmetry, supergravity, superstrings, and recently
branes technology. Application of octonions to supergravity spontaneous compactification was a
very important and active field of research during the mid-1980s, especially compactification
of $d=11$ supergravity over $\mathbb S^7$ to four dimensions. It is an impossible task to list
all the relevant papers, so we direct the interested reader to Ref.~2, there a lot of
references are given. We just mention that the first indication of the octonionic nature of
this problem appeared in the Englert solution of $d=11$ supergravity over $\mathbb S^7$. The
relation between superstrings ($p$ branes) and octonions had been considered from many
different points of view, the reader may consult the references given in~Ref.~3 for details.
Recently nonassociativity is known to appear in open string theory with nonconstant background
$B_{\mu\nu}$ field, see Ref.~4. It was also argued that the algebra of closed string field
theory should be commutative nonassociative. In~Ref.~5, they discussed commutative
nonassociative gauge theory with Lorentz and Poincar\'e symmetry. There also other discussions
on nonassociative theory, see Ref.~6.
\par
In the paper we attempt to construct a gauge theory based on the octonion algebra in familiar
manner of Yang-Mills theory. The paper is organized as follows. In Section 2 we list the
properties of octonions and some other mathematical structures relevant to our work. In
Section 3 we extend the gauge invariance to the theory in which a nonassociative Moufang loop
take the place of the structure group. In Section 4 we construct an on-shell version of $N=1$
supersymmetric gauge theory without matter. Section 5 contain some concluding remarks. In
order to make the paper self-consistent we present in appendix the some useful formula for
gamma matrices and Majorana spinors.

\section{Notations and preliminary results}

To define our notations we list the features of octonion algebra and some other mathematical
structures as far as they are of relevance to our work. In addition, we prove some simple
assertions concerning isomorphisms and automorphisms of the octonion algebra.

\subsection{Algebra of octonions}

We recall that the algebra $\mathbb O$ of octonions is a real linear algebra with the
canonical basis $1,e_{1},\dots,e_{7}$ such that
\begin{equation}
e_{i}e_{j}=-\delta_{ij}+c_{ijk}e_{k},
\end{equation}
where the structure constants $c_{ijk}$
are completely antisymmetric and nonzero and equal to unity for the seven combinations (or
cycles)
$$
(ijk)=(123),(145),(167),(246),(275),(374),(365).
$$
The algebra of octonions is not associative but alternative, i.e. the associator
\begin{equation}
(x,y,z)=(xy)z-x(yz)
\end{equation}
is totally antisymmetric in $x,y,z$. Consequently, any two elements of $\mathbb O$ generate an
associative subalgebra. The algebra $\mathbb O$ permits the involution (antiautomorphism of
period two) $x\to\bar x$ such that the elements
\begin{equation}
t(x)=x+\bar x\qquad\text{and}\qquad n(x)=\bar xx
\end{equation}
are in $\mathbb R$. In the canonical basis this involution is defined by $\bar e_{i}=-e_{i}$.
It follows that the bilinear form
\begin{equation}
\langle x,y\rangle=\frac12(\bar xy+\bar yx)
\end{equation}
is positive definite and defines an inner product on $\mathbb O$. Obviously, it is invariant
under all automorphisms of $\mathbb O$. It is easy to prove that the quadratic form $n(x)$ is
positive definite and permits the composition
\begin{equation}
n(xy)=n(x)n(y).
\end{equation}
Since the quadratic form $n(x)$ is positive definite, it follows immediately from (2.5) that
$\mathbb O$ is a division algebra.
\par
There is an explicit procedure for building the algebra of octonions. Suppose $e$ is an
element in $\mathbb O$ such that $\bar e=-e$ and $n(e)=1$. We choose a quaternion subalgebra
$\mathbb H$ so that $\mathbb H\perp e$ and define a multiplication on the vector space direct
sum $\mathbb H\oplus\mathbb He$ by
\begin{equation}
(x_1+y_1e)(x_2+y_2e)=(x_1x_2-\bar y_2y_1)+(y_2x_1+y_1\bar x_2)e.
\end{equation}
Obviously, $\mathbb He$ is an orthogonal complement to $\mathbb H$ relative to the form (2.4).
We denote this space by the symbol $\mathbb H^{\perp}$. It can easily be checked that the
algebra $\mathbb H\oplus\mathbb H^{\perp}$ with the multiplication (2.6) is the algebra of
octonions. Note also that $\mathbb O$ is unique, to within an isomorphism, nonassociative
composition division algebra. We refer for proof to Ref.~7.

\subsection{Malcev algebras and Moufang loops}

Since the algebra of octonions is nonassociative, its commutator algebra $\mathbb O^{(-)}$ is
non-Lie. Instead of the Jacobi identity the algebra $\mathbb O^{(-)}$ satisfies the Malcev
identity
\begin{equation}
J(x,y,[x,z])=[J(x,y,z),x],
\end{equation}
where
\begin{equation}
J(x,y,z)=[[x,y],z]+[[y,z],x]+[[z,x],y]
\end{equation}
is so-called Jacobian of $x,y,z$. We define the seven-dimensional Malcev subalgebra
\begin{equation}
\mathbb M=\{x\in\mathbb O^{(-)}\mid t(x)=0\}.
\end{equation}
Obviously, the algebra $\mathbb M$ has the basis $e_{1},\dots,e_{7}$. Using (2.1) we can find
the commutators and Jacobians of the basis elements
\begin{align}
[e_{i},e_{j}]&=2c_{ijk}e_{k},\\
J(e_{i},e_{j},e_{k})&=12c_{ijkl}e_{l}.
\end{align}
Here $c_{ijkl}$ is a completely antisymmetric nonzero tensor equal to unity for the seven
combinations
$$
(ijkl)=(4567),(2367),(2345),(1357),(1364),(1265),(1274).
$$
An anticommutative algebra satisfying the identity (2.7) is called a Malcev algebra$^8$. The
Malcev algebra (2.9) has a particular importance. It is known$^9$ that any real compact simple
non-Lie Malcev algebra is isomorphic to the algebra $\mathbb M$. In addition, any semisimple
Malcev algebra of characteristic 0 is decomposed in a direct sum of simple algebras. In
particular, any semisimple Malcev algebra is isomorphic to a subalgebra of commutator algebra
for some alternative algebra.
\par
Recall$^{10}$, that a loop is a binary system with a unity element, in which the equations
$ax=b$ and $ya=b$ are uniquely solvable. An analytic loop is an analytic manifold equipped
with the loop structure, in which the binary operations are analytic. Since the algebra of
octonions $\mathbb O$ is a real division algebra, the set $\mathbb O^{*}$ of all nonzero
elements of $\mathbb O$ is an analytic loop. It is easy to prove that $\mathbb O^{*}$
satisfies the identities
\begin{equation}
(xy)(zx)=x((yz)x),\quad ((zy)z)x=z(y(zx)),\quad x(y(zy))=((xy)z)y
\end{equation}
which are called the central, left, and right Moufang identities accordingly. In any loop two
of them are a corollary of third. A loop is called a Moufang loop if it satisfies the
identities (2.12). Note that by (2.5) the set
\begin{equation}
\mathbb S=\{x\in\mathbb O^{*}\mid n(x)=1\}
\end{equation}
is closed relative to the multiplication defined by (2.6). Consequently, $\mathbb S$ is an
analytic Moufang loop. It is known$^{11}$ that $\mathbb S$ is unique, to within an
isomorphism, analytic compact simple nonassociative Moufang loop and its tangent algebra is
isomorphic to the Malcev algebra $\mathbb M$. In addition, any semisimple analytic Moufang
loop is decomposed in a direct product of simple Moufang loops. Everywhere below we denote the
algebra (2.9) and the loop (2.13) by the symbols $\mathbb M$ and $\mathbb S$ respectively.

\subsection{Isomorphisms and automorphisms}

We will use the following construction. Let $u$ be a fixed element of $\mathbb S$. We define a
new multiplication in $\mathbb O$ by
\begin{equation}
x\circ y=(xu^{-3})(u^{3}y).
\end{equation}
Obviously, the multiplication (2.14) converts the vector space $\mathbb O$ into a linear
algebra. We denote this algebra by the symbol $\mathbb O'$. It is easy to prove the following.
\medskip
\par\noindent
{\bf Proposition 1.} {\it The algebras $\mathbb O$ and $\mathbb O'$ are isomorphic.}
\medskip
\par\noindent
{\it Proof.} In the first place, note that the algebra $\mathbb O'$ is composition. Indeed,
the quadratic form $n(x)=\bar xx$ is defined on the space $\mathbb O'$. Using (2.5), we prove
the identity $n(x\circ y)=n(x)n(y)$. Secondly, the equations $a\circ x=b$ and $y\circ a=b$ are
uniquely solvable in $\mathbb O'$. In the third place, the dimensions of $\mathbb O$ and
$\mathbb O'$ are coincided. Thus, $\mathbb O'$ is an eight-dimensional composition division
algebra. Using the classification of composition algebras, we prove the isomorphism $\mathbb
O'\simeq\mathbb O$.
\medskip\par
We construct the isomorphism $\mathbb O\to\mathbb O'$ in the explicit form. Supposed $\mathbb
H$ is a quaternion subalgebra in $\mathbb O$ such that $u\in\mathbb H$. We consider the
mapping $\alpha:\mathbb O\to\mathbb O'$ such that
\begin{equation}
\begin{aligned}
\alpha(x)&=x,&\text{if}\quad x&\in\mathbb H,\\
\alpha(x)&=u^{-3}x,\qquad&\text{if}\quad x&\in\mathbb H^{\perp}.
\end{aligned}
\end{equation}
\par\noindent
{\bf Proposition 2.} {\it The mapping  $\alpha:\mathbb O\to\mathbb O'$ defined by (2.15) is an
isomorphism of the algebras.}
\medskip
\par\noindent
{\it Proof.} Denote by $x'$ the element $\alpha(x)$. Using (2.6) we proof by direct
calculation that
\begin{align}
(xy)'&=x'\circ y',\notag\\
(x(ye))'&=x'\circ(ye)',\notag\\
((ye)x)'&=(ye)'\circ x',\notag\\
((xe)(ye))'&=(xe)'\circ(ye)'\notag
\end{align}
for any $x,y\in\mathbb H$ and $e\in\mathbb H^{\perp}$ with $n(e)=1$. Consequently, the mapping
$\alpha:\mathbb O\to\mathbb O'$ is an isomorphism.
\medskip\par
The equalities (2.15) define not only the isomorphism $\mathbb O\to\mathbb O'$ of the algebras
but also a linear transformation of the space $\mathbb O$. Suppose
\begin{align}
\beta(x)&=uxu^{-1},\\
\varphi(x)&=ux'u^{-1}.
\end{align}
[Here and everywhere below we denote the element $\alpha(x)$ by the symbols $x'$.] Then we
have the following.
\medskip
\par\noindent
{\bf Proposition 3.} {\it The linear transformation $\varphi=\beta\alpha$, defined by (2.17)
is an automorphism of $\mathbb O$.}
\medskip
\par\noindent
{\it Proof.} On the one hand,
\begin{equation}\begin{split}
\varphi(xy)&=u(x'\circ y')u^{-1},\\
\varphi(x)\varphi(y)&=(ux'u^{-1})(uy'u^{-1}).
\end{split}\end{equation}
On the other hand, it follows from the Moufang identities (2.12) that
\begin{equation}
x^{-1}((xy)z)=(yx^{-1})(xz)=(y(zx^{-1}))x.
\end{equation}
Using alternativity of $\mathbb O$ and the identities (2.19), we get
\begin{equation}
u(x\circ y)u^{-1}=(uxu^{-1})(uyu^{-1}).
\end{equation}
Comparing (2.18) and (2.20), we prove the proposition.
\medskip\par
Obviously, the mapping (2.15) is not defined only by selection of $u$. We must also fix a
quaternion subalgebra of $\mathbb O$ containing this element. To this end we fix an element
$\psi$ in $\mathbb M$. It is obvious that the coupe $(u,\psi)$ generates a quaternion
subalgebra if $u\psi u^{-1}\ne \psi$. In this case we say that $(u,\psi)$ defines the
transformations (2.15) and (2.17).
\medskip
\par\noindent
{\bf Proposition 4.} {\it Let $\psi$ be a fix element of $\mathbb M$. Then the transformations
(2.17) defined by $(u,\psi)$ for all $u\in\mathbb S$ generate the groups} $\text{Aut}\,\mathbb
O$ {\it of all automorphisms of $\mathbb O$.}
\medskip
\par\noindent
{\it Proof.} First note that all such transformations generate a subgroup of
$\text{Aut}\,\mathbb O$. Further, let $x$ be a nonzero element in $\mathbb M$ and $\mathbb H$
be a quaternion subalgebra of $\mathbb O$ containing $x$ and $\psi$. Then for any $u$ in
$\mathbb H$ such that $uxu^{-1}\ne x$ the transformation defined by $(u,\psi)$ does not leave
fixed $x$. On the other hand, the group $\text{Aut}\,\mathbb O$ is isomorphic to $G_2$.
Therefore any maximum subgroup of $\text{Aut}\,\mathbb O$ is isomorphic either to $SU(3)$ or
$SO(4)$. If we observe that these subgroups leave fixed the elements of $\mathbb M$, we prove
the proposition.

\section{Nonassociative gauge theory}

In this section we construct a nonassociative gauge theory. At first we give a brief summary
of representation theory of Malcev algebras. Then we introduce gauge fields taking their
values in the algebra $\mathbb M$ and find a transformation law of these fields under the
gauge transformations. Further, we construct a field strength tensor and find its
transformation law under these transformations. In the end of the section we show that our
theory admits the Hamilton gauge.

\subsection{Representations of Malcev algebras}

Let $M$ be a finite-dimensional semisimple Malcev algebra over a field $F$ of characteristic
0. Without loss of generality it can be assumed that the algebra $M$ is embedded is a
commutator algebra of alternative algebra. Suppose $V$ is a vector space over $F$ and
$T:M\to\text{End\,}V$ $(x\to T_{x})$ is a liner mapping. Then $T$ is called a representation
of $M$ if this algebra defined on the direct sum $M\oplus V$ by means of
\begin{equation}
[v+x,w+y]=T_{x}w-T_{y}v+[x,y]
\end{equation}
is a Malcev algebra. In this case $V$ is said to be a Malcev module for $M$ or $M$ module. It
follows from (2.7) that the operators $T_{x}$ satisfy
\begin{equation}
T_{[[x,y],z]}=T_{z}T_{y}T_{x}-T_{y}T_{x}T_{z}+T_{x}T_{[y,z]}-T_{[z,x]}T_{y}.
\end{equation}
Conversely, if for all $x,y,z\in M$ the equation (3.2) is true, then $T$ is a representation
of $M$.
\par
A special case of the representation is the mapping $T:M\to\text{End\,}M$ that defined by the
equations
\begin{equation}
T_{x}y=[x,y]
\end{equation}
for all $y\in M$. This representation is said to be regular (or adjoint). Second example of
the representation comes out if we consider the mapping $T:M\to\text{End\,}V$ satisfying
\begin{equation}
T_{[x,y]}=[T_{x},T_{y}]
\end{equation}
for all $x,y\in M$. Since (3.2) is a corollary of (3.4), this mapping is really
representations of $M$ (and a homomorphism of $M$ into a Lie algebra of linear transformations
of $V$). Such representations are important for the theory of Lie algebras; however, their
significance is not too large in the theory of Malcev algebras.
\par
Nevertheless, the representation theory of Malcev algebras is analogous to the representation
theory of Lie algebras. It is  known$^9$ that any representation of a semisimple Malcev
algebra is completely reducible. Any irreducible Malcev module is either Lie or the regular
module for a nonassociative simple Malcev algebra or $sl(2)$ module of dimension 2 such that
$T_{x}=x^{*}$, where $x^{*}$ is the adjoint matrix to $x\in sl(2)$. Note also that the
representation theory can be extend to Moufang loops$^{12}$.
\par
The situation is very simple if we have the algebra $\mathbb M$. Any nontrivial representation
of $\mathbb M$ is regular; the operators $T_{x}$ are defined by (3.3) and generate the Lie
algebra $so(7)$. The latter is decomposed into the direct sum $D(\mathbb M)\oplus T(\mathbb
M)$ of the algebra $D(\mathbb M)$ of derivations of $\mathbb M$ and the seven-dimensional
subspace $T(\mathbb M)$. In addition, the Lie brackets are given by
\begin{align}
[T_{x},T_{y}]&=D_{x,y}-T_{[x,y]},\\
[D_{x,y},T_{z}]&=T_{D_{x,y}z},
\end{align}
where $D_{x,y}$ is an operator of derivations of $\mathbb M$. It is well known that the
algebra $D(\mathbb M)$ is isomorphic to the exceptional Lie algebra $g_2$. Obviously, the
algebras of derivations of $\mathbb M$ and $\mathbb O$ are coincided.

\subsection{Gauge transformations}

We will now apply the representation theory of Malcev algebras to a construction of gauge
theory. Let $A_{\mu}(x)$ be a vector field taking its value in $\mathbb M$ and $\psi(x)$ be a
field taking its value in a space $V$ of representation of $\mathbb M$. Denote by $\hat A_{\mu
}$ the operator $T_{A_{\mu}}$ and define the covariant derivative
\begin{equation}
D_{\mu}\psi=\partial_{\mu}\psi+\hat A_{\mu}\psi.
\end{equation}
Obviously, the spaces $V$ and $\mathbb M$ are coincided and the operator $\hat A_{\mu}$ is
defined by
\begin{equation}
\hat A_{\mu}\psi=[A_{\mu},\psi].
\end{equation}
As in the Yang-Mills theory, the gauge field is endowed with a transformation law under gauge
transformations such that $D_{\mu}\psi$ transform as $\psi$, i.e.,
\begin{align}
\psi&\to U\psi,\\
D_{\mu}\psi&\to U(D_{\mu}\psi),
\end{align}
where $U=U(x)$ is  a function taking its values in the group $\text{Aut\,}M$ of all
automorphisms of $\mathbb M$.
\par
We will now find a transformation law of $A_{\mu}$ under the gauge transformations (3.9). From
(3.9) and (3.10), we get the usual transformation law of operator functions
\begin{equation}
\partial_{\mu}+\hat A_{\mu}\to\partial_{\mu}+U\hat A_{\mu}U^{-1}
+U\partial_{\mu}U^{-1}.
\end{equation}
Since $\hat A_{\mu}=T_{A_{\mu}}$ and $U\in \text{Aut\,}M$, we have
\begin{equation}
U\hat A_{\mu}U^{-1}=T_{UA_{\mu}}.
\end{equation}
On the other hand, it follows from Propositions 3 and 4 that the function $U(x)$ can be chosen
as the composition $U=\beta\alpha$ of transformations defined by (2.15) and (2.16). By
Proposition 2, it follows that the operator function $\alpha(x)$ defines the isomorphism
$\mathbb O\to\mathbb O'$ for any value of $x$. Suppose $\psi'=\alpha(\psi)$ and define its
derivation by
\begin{equation}
\nabla_{\mu}\psi'=(\partial_{\mu}\psi)'.
\end{equation}
It is easy to prove that any two differentiable functions $f(x)$ and $g(x)$ taking their
values in $\mathbb O'$ satisfy
\begin{equation}
\nabla_{\mu}(f\circ g)=\nabla_{\mu}f\circ g +f\circ\nabla_{\mu}g.
\end{equation}
Noting that the operator
\begin{equation}
\nabla_{\mu}=\partial_{\mu}+\alpha\partial_{\mu}\alpha^{-1},
\end{equation}
and using (3.12) and (3.14), we get
\begin{equation}
\partial_{\mu}+T_{A_{\mu}}\to\nabla_{\mu}+T_{UA_{\mu}}+\beta\nabla_{\mu}\beta^{-1}
\end{equation}
instead of (3.11).
\par
Suppose that the transformations (3.9) and (3.10) are infinitesimal. Then the operator
functions $\alpha$ and $\beta$ take the form
\begin{align}
\alpha(x)&=1+\varGamma(x)\\
\beta(x)&=1+T_{\theta(x)},
\end{align}
where $\theta(x)$ is defined by $u(x)=1+\theta(x)$ that takes its value in a neighborhood of
unity element of $\mathbb S$. In this case we can consider the transformations
\begin{align}
\partial_{\mu}&\to\nabla_{\mu}=\partial_{\mu}-\partial_{\mu}\varGamma,\\
A_{\mu}&\to A'_{\mu}+[\theta,A_{\mu}]-\partial_{\mu}\theta,
\end{align}
where $A'_{\mu}=A_{\mu}+\varGamma A_{\mu}$, instead of (3.16). The formula (3.20) gives us a
transformation law of $A_{\mu}$ under the gauge transformations (3.9). Notice that in contrast
with the Yang-Mills theory, we have the transformation (3.19). As usual, we define a finite
gauge transformation as an infinite sequence of infinitesimal transformations.
\par
We now want to construct the field strength tensor in the nonassociative case. Denote by $\hat
F_{\mu\nu}$ a projection of the Lie bracket $[D_{\mu},D_{\nu}]$ onto $T(\mathbb M)$. Using
(3.5), we get
\begin{equation}
\hat F_{\mu\nu}=T_{F_{\mu\nu}},
\end{equation}
where the tensor
\begin{equation}
F_{\mu\nu}=\partial_{\mu}A_{\nu}-\partial_{\nu}A_{\mu}-[A_{\mu},A_{\nu}].
\end{equation}
Since the subspace $T(\mathbb M)$ is $G_2$ invariant, it follows that (3.11) induces the
transformation
\begin{equation}
\hat F_{\mu\nu}\to U\hat F_{\mu\nu}U^{-1}.
\end{equation}
Using (3.12) and (3.21), we get the transformation law
\begin{equation}
F_{\mu\nu}\to UF_{\mu\nu}=F'_{\mu\nu}+[\theta,F_{\mu\nu}]
\end{equation}
of the tensor (3.22) under the infinitesimal gauge transformations (3.19) and (3.20). It
follows from (3.24) that the field strength tensor may be really defined by (3.22). Notice
that the tensor $F_{\mu\nu}$ takes more habitual form in the basis $\tilde e_{i}=-e_{i}$. In
this basis $\tilde A_{\mu}=-A_{\mu}$ and $\tilde F_{\mu\nu}=-F_{\mu\nu}$.

\subsection{Hamilton gauge}

In the Yang-Mills theory, owing to the gauge arbitrariness, we may demand that the potential
locally satisfies a definite condition. The situation is similar in the nonassociative case.
There exists a gauge transformation $A_{\mu}\to A_{\mu}^{\varphi}$ such that
\begin{equation}
A_0^{\varphi}(x)=0.
\end{equation}
Indeed, the potential $A_0(\mathbf x,t)\to 0$ as $t\to-\infty$. Therefore there exists $t_1$
such that the equation
\begin{equation}
\frac{\partial u}{\partial t}=uA_0,
\end{equation}
where $u(x)$ takes its values in a neighborhood of unity element of $\mathbb S$, has the
solution
\begin{equation}
u_1(\mathbf x,t)=1+\int^{t}_{-\infty} A_0(\mathbf x,s)ds\notag
\end{equation}
for all $t\in[-\infty,t_1]$. Since the mapping $\mathbb O\to\mathbb O'$ defined by (2.15) is
isomorphism, it follows from (3.26) that the function
\begin{equation}
A_{0,1}=u_1(A'_0+\nabla_0)u_1^{-1}\notag
\end{equation}
satisfies $A_{0,1}=0$ on this interval. It is clear that $A_0$ and $A_{0,1}$ are connected by
an infinitesimal gauge transformation.
\par
Further, let $t_{n+1}=t_{n}+\delta t_{n}$, where $n\in\mathbb N$. It is readily seen that the
equation
\begin{equation}
\frac{\partial u}{\partial t}=uA_{0,n}
\end{equation}
has the solution
\begin{equation}
u_{n+1}(\mathbf x,t)=1+\int^{t}_{-\infty} A_{0,n}(\mathbf x,s)ds\notag
\end{equation}
on the interval $[-\infty,t_{n+1}]$ if the function $A_{0,n}(x)$ is defined by
\begin{equation}
A_{0,n}=u_{n}\left(A'_{0,n-1}+\nabla_0\right)u_{n}^{-1}\notag
\end{equation}
with $A_{0,0}=A_0$. From (3.27) it follows that $A_{0,n+1}=0$ for all $t<t_{n+1}$. If we
suppose
\begin{equation}
A_{0}^{\varphi}(x)=\lim_{n\to\infty}A_{0,n}(x),\notag
\end{equation}
and use the induction on $n$, we prove (3.25). The functions $A_0(x)$ and $A_{0}^{\varphi}(x)$
are connected by a gauge transformation. Hence in every class of gauge-equivalent fields,
there exists a field satisfying the condition (3.25).

\section{Supersymmetric gauge theory}

In this section we construct an on-shell version of $N=1$ supersymmetric gauge theory without
matter. The model is described by a vector field $A_{\mu}$ and by a Majorana spinor field
$\psi$. All fields take their values in the Malcev algebra $\mathbb M$.

\subsection{Supersymmetry transformations}

We examine the Lagrangian density
\begin{equation}
\mathcal L=-\frac{1}{4}\langle F_{\mu\nu},F^{\mu\nu}\rangle
+\frac{i}{2}\langle\bar\psi,\gamma^{\mu}D_{\mu}\psi\rangle.
\end{equation}
It contains the covariant derivative $D_{\mu}\psi$ and the field strength tensor $F_{\mu\nu}$
defined by (3.7) and (3.22), respectively. Since the inner product (2.4) is invariant under
all automorphisms of $\mathbb O$, it follows from (3.9), (3.10) and (3.24) that the Lagrangian
density (4.1) is invariant under the gauge transformations (3.19) and (3.20).
\par
We will prove that the action with this Lagrangian density is invariant under the following
supersymmetry transformations:
\begin{align}
\delta A_{\mu}&= i\bar\varepsilon\gamma_{\mu}\psi,\\
\delta \psi&= \frac12F_{\mu\nu}\gamma^{\mu\nu}\varepsilon,
\end{align}
where $\varepsilon$ is a constant anticommuting Majorana spinor. To calculate the variation of
the Lagrangian density one needs the formulas (A.2), (A.3), (A.6) and (A.8) in the Appendix.
Using these formulas and the identities
\begin{align}
\delta F_{\mu\nu} &= i\bar\varepsilon(\gamma_{\nu}D_{\mu}-\gamma_{\mu}D_{\nu})\psi,\notag\\
\delta(D_{\mu}\psi) &= D_{\mu}\delta\psi+[\delta A_{\mu},\psi],\notag
\end{align}
we get
\begin{equation}
\delta\mathcal L
=\frac{1}{2}\langle\bar\psi\gamma^{\mu},[\psi,\bar\varepsilon\gamma_{\mu}\psi]\rangle
+\frac{i}{2}\langle D_{\rho}F_{\mu\nu}, \bar\varepsilon\gamma^{\mu\nu\rho}\psi\rangle
-\frac{1}{2}\bar\varepsilon\partial_{\mu}V^{\mu},
\end{equation}
with
\begin{equation}
V^{\mu}=\langle iF^{\mu\nu},\gamma_{\nu}\psi\rangle+\langle
^*\!F^{\mu\nu},\gamma_5\gamma_{\nu}\psi\rangle,\notag
\end{equation}
where $^{*}\!F^{\mu\nu}$ is a dual field strength tensor.
\par
It is easy to prove that the first term in the right hand side of (4.4) vanishes. Indeed, the
tensor $c_{ijk}$ defined by (2.1) is completely antisymmetric. Therefore we can act as in the
supersymmetric Yang-Mills theory. We rewrite this term as
\begin{equation}
\frac12\langle\bar\psi\gamma^{\mu},[\psi,\bar\varepsilon\gamma_{\mu}\psi]\rangle
=c_{ijk}(\bar\varepsilon\gamma_{\mu}\psi^{k})(\bar\psi^{i}\gamma^{\mu}\psi^{j}).
\end{equation}
Then we insert the Fierz identity (A.5) for $\psi^{k}\bar\psi^{i}$ in the right hand side of
(4.5) and use the relations (A.4) in the appendix. We get
\begin{multline} (\bar\psi^{i}\gamma^{\mu}\psi^{j})(\bar\varepsilon\gamma_{\mu}\psi^{k})
=-(\bar\varepsilon\psi^{j})(\bar\psi^{i}\psi^{k})
+\frac12(\bar\varepsilon\gamma_{\mu}\psi^{j})(\bar\psi^{i}\gamma^{\mu}\psi^{k})\\
-\frac12(\bar\varepsilon\gamma_{\mu}\gamma_{5}\psi^{j})
(\bar\psi^{i}\gamma_{5}\gamma^{\mu}\psi^{k})
+(\bar\varepsilon\gamma_{5}\psi^{k})(\bar\psi^{i}\gamma_{5}\psi^{k}),\notag
\end{multline}
where all but the second term on the right hand side is symmetric in $i$ and $k$. Using this
identity, we prove that the expression on the left in (4.5) is zero.
\par
We now examine the second term in the right hand side of (4.4). Since the algebra $\mathbb M$
is non-Lie, the tensor $^{*}\!F_{\mu\nu}$ does not satisfy the Bianchi identity. Therefore it
is not obvious that this term is zero. Let $\eta$ be a constant anticommuting Majorana spinor
such that $\bar\eta\eta=1$, and let $\varepsilon=a\eta$ for $a\in\mathbb R$. Using the
identities (A.9) and (A.10) in the Appendix, we get
\begin{equation}
(\bar\eta\gamma_5\eta)(\bar\varepsilon\gamma_{\mu}\psi)
=\bar\psi\gamma_{\mu}\gamma_5\varepsilon.
\end{equation}
Using (4.6) and (A.3), we get
\begin{equation}
k\varepsilon^{\mu\nu\rho\sigma}(\bar\varepsilon\gamma_{\sigma}\psi)
=\bar\varepsilon\gamma^{\mu\nu\rho}\psi,
\end{equation}
where $k=i\bar\eta\gamma_5\eta$. It follows from (4.7) that
\begin{equation}
i\langle D_{\rho}F_{\mu\nu},
\bar\varepsilon\gamma^{\mu\nu\rho}\psi\rangle=k\varepsilon^{\mu\nu\rho\sigma}\langle
D_{\rho}F_{\mu\nu},i\bar\varepsilon \gamma_{\sigma}\psi\rangle.
\end{equation}
On the other hand, it is easy to prove that
\begin{equation}
3\varepsilon^{\mu\nu\rho\sigma}D_{\rho}F_{\mu\nu}
=\varepsilon^{\mu\nu\rho\sigma}J(A_{\mu},A_{\nu},A_{\rho}),
\end{equation}
where the Jacobian $J(A_{\mu},A_{\nu},A_{\rho})$ is defined by (2.8). Using (2.11) and (4.2),
we get
\begin{equation}
\varepsilon^{\mu\nu\rho\sigma}\langle D_{\rho}F_{\mu\nu},i\bar\varepsilon
\gamma_{\sigma}\psi\rangle=\varepsilon^{\mu\nu\rho\sigma}c_{ijkl} \delta
(A^{i}_{\mu}A^{j}_{\nu}A^{k}_{\rho}A^{l}_{\sigma}).
\end{equation}
In the Hamilton gauge the right hand side of (4.10) vanishes. Since the action with the
Lagrangian density defined in (4.1) is gauge invariant, we conclude that the second term in
(4.4) is absent and that the supersymmetric variation of the Lagrangian density is just a
divergence.

\subsection{Superalgebra}

A basic algebraic fact about supersymmetry is that the commutator of two supersymmetry
transformations gives a spatial translation. This is true for our theory. Indeed, using the
formulas (A.1), (A.2), (A.6), and (A.7) in the Appendix and the obvious identity
$\gamma_{\mu\nu}\gamma^{\nu}=3\gamma_{\mu}$, we prove that on shell the commutators
\begin{align}
[\delta_1,\delta_2]A_{\mu}&
=-2i(\bar\varepsilon_2\gamma^{\nu}\varepsilon_1)\partial_{\nu}A_{\mu}+D_{\mu}\theta,\\
[\delta_1,\delta_2]\psi&
=-2i(\bar\varepsilon_2\gamma^{\nu}\varepsilon_1)\partial_{\nu}\psi+[\psi,\theta].
\end{align}
The gauge parameter $\theta=2i(\bar\varepsilon_2\gamma^{\nu}A_{\nu} \varepsilon_1)$ depends on
the gauge field and the supersymmetric parameters $\varepsilon_{i}$. Here we use the fact that
$\psi$ obeys the Dirac equation $\gamma^{\mu}D_{\mu}\psi=0$. Further, we consider the
consequence
\begin{equation}
(A_{\mu},\psi)\overset U\to(\tilde A_{\mu},\tilde\psi)\overset\varPhi\to(\Tilde{\Tilde
A}_{\mu},\Tilde{\Tilde\psi})\notag
\end{equation}
of two gauge transformations $U$ and $\varPhi$. Here $U$ is an infinitesimal transformation
and $\varPhi$ is a finite transformation. It follows from (3.9) and (3.20) that the
transformation $U$ is
\begin{align}
\delta\psi&=\varGamma\psi+[\theta,\psi],\notag\\
\delta A_{\mu}&=\varGamma A_{\mu}+[\theta,A_{\mu}]-\partial_{\mu}\theta.\notag
\end{align}
On the other hand, it follows from (3.19) that the transformation $\varPhi$ defines the
mapping $\partial_{\mu}\to\nabla_{\mu}$, where the covariant operator $\nabla_{\mu}$ is given
by (3.15). If we choose the infinitesimal function
\begin{equation}
\varGamma=-2i(\bar\varepsilon_2\gamma^{\nu}\varepsilon_1)\alpha\partial_{\mu}\alpha^{-1},\notag
\end{equation}
then from (4.11) and (4.12) we get the operator relation
\begin{equation}
[\delta_1,\delta_2] =-2i(\bar\varepsilon_2\gamma^{\nu}\varepsilon_1)\partial_{\nu}.
\end{equation}
Thus, as in the supersymmetric Yang-Mills theory this superalgebra closes only on gauge
invariant fields.

\subsection{Chiral representation}

In spite of the fact that in the simplest $N=1$ supersymmetric gauge theories one usually uses
Majorana spinors, it is very desirable to examine our pattern in the chiral representation.
Primarily, we rewrite the Lagrangian density and the supersymmetry transformations of the
theory in terms of Weyl spinors. Suppose
\begin{equation}
\mathcal L'=-\frac{1}{4}\langle F_{\mu\nu},F^{\mu\nu}\rangle
+\frac{i}{2}\langle\bar\psi,\gamma^{\mu}D_{\mu}\psi\rangle+\frac{k}{4}\langle
F_{\mu\nu},^{*}\!F^{\mu\nu}\rangle,
\end{equation}
where $\psi$ is a left-handed spinor and $k$ is a constant. It is obvious that $\mathcal L'$
is invariant under the gauge transformations (3.19) and (3.20). We consider the following
supersymmetry transformations:
\begin{align}
\delta A_{\mu}&=\frac{i}{2}\left\{\bar\varepsilon\gamma_{\mu}\psi
-(\bar\varepsilon\gamma_{\mu}\psi)^{\dag}\right\},\\
\delta \psi&= \frac12F_{\mu\nu}\gamma^{\mu\nu}\varepsilon,
\end{align}
where $\varepsilon$ is a constant anticommuting left-handed Weyl spinor. As above, we
calculate the variation
\begin{equation}
\delta\mathcal L' =\frac{i}{2}\langle D_{\rho}F_{\mu\nu},
(\bar\varepsilon\gamma^{\mu\nu\rho}\psi
-k\varepsilon^{\mu\nu\rho\sigma}\bar\varepsilon\gamma_{\sigma}\psi)\rangle
-\frac{1}{2}\bar\varepsilon\partial_{\mu}\widetilde V^{\mu}+H.c.
\end{equation}
Using the formula $\gamma_5\psi=\psi$ and the identities (A.3) in the Appendix, we prove that
the first term in the right hand side of (4.17) vanishes only if $k=i$. Thus, the last term in
the right hand side of (4.14) is purely imaginary. Arguing as the end of Subsection 4.1, we
see that in four dimensions the last term in the right hand side of (4.14) is a divergence
though. Consequently, the action with the Lagrangian density defined in (4.14) is invariant
under the supersymmetry transformations (4.15) and (4.16).

\section{Conclusion}

In this paper we have given a construction of nonassociative gauge theory in which the Moufang
loop is used instead of the structure group. We have also demonstrated how this theory can be
used to construct an on-shell version of $N=1$ supersymmetric gauge theory without matter.
\par
In contrast to the Yang-Mills theory, we have studied not only transformations of the gauge
field but also transformations of the operator of differentiation. This is a characteristic
feature of the gauge theory. Because of this we may demand that the potential locally
satisfies a definite condition. In particular, we may choose the Hamilton gauge. It is obvious
that the gauge theory can be defined in spaces of dimension greater than 4. In addition, it
can be easily generalized if we take a real semisimple Malcev algebra instead of the algebra
$\mathbb M$. Since any Lie algebra is Malcev, it follows that such gauge theory is a
generalization of the Yang-Mills theory.
\par
Conversely, it is not clear how the $N=1$ supersymmetric gauge theory can be defined in spaces
of dimension greater than 4 and how the simplest four-dimensional supersymmetric theory can be
extended to theories with extended supersymmetry. In addition, there is the challenge to
couple the $N=1$ supersymmetric gauge theory to the supergravity system so that the combined
system is invariant under the local supersymmetric transformations. It is unsatisfying to be
limited to the simple example of supersymmetric gauge theory that we have considered without
evidence that more general possibilities are not viable. Therefore there are a lot of open
problems, which deserve further study.

\bigskip\medskip\par\noindent
{\bf Acknowledgements}
\medskip\par\noindent
The research was supported by RFBR Grant No. 06-02-16140.

\appendix
\section{Appendix}

In this appendix we collected some useful formulas which are used in the main body of the
paper. In particular, we use the commutation relations
\begin{equation}
[\gamma_{\mu\nu},\gamma_{\rho\sigma}]
=2(\gamma_{\mu\sigma}g_{\nu\rho}+\gamma_{\nu\rho}g_{\mu\sigma}
-\gamma_{\mu\rho}g_{\nu\sigma}-\gamma_{\nu\sigma}g_{\mu\rho}),
\end{equation}
of gamma matrices in four dimensions and the products
\begin{equation}\begin{split}
\gamma_{\mu\nu}\gamma_{\rho}&=\gamma_{\mu\nu\rho}+g_{\nu\rho}\gamma_{\mu}
-g_{\mu\rho}\gamma_{\nu},\\
\gamma_{\rho}\gamma_{\mu\nu}&=\gamma_{\mu\nu\rho}-g_{\nu\rho}\gamma_{\mu}
+g_{\mu\rho}\gamma_{\nu},
\end{split}\end{equation}
where we use the notation $\gamma_{\mu\nu\dots}$ for a totally antisymmetrized product of
$\gamma_{\mu}\gamma_{\nu}\dots$. We also use the simple relations
\begin{equation}\begin{split}
\gamma^{\mu\nu\rho\sigma}\gamma_{\sigma}&=\gamma^{\mu\nu\rho},\\
\varepsilon^{\mu\nu\rho\sigma}\gamma_5&=i\gamma^{\mu\nu\rho\sigma},
\end{split}\end{equation}
where $\gamma_{5}=-i\gamma_{0}\gamma_{1}\gamma_{2}\gamma_{3}$, and the conjugation formulas
\begin{align}
\gamma_{\mu}\gamma_{\rho}\gamma^{\mu} &= -2\gamma_{\rho},\notag\\
\gamma_{\mu}\gamma_{\rho\sigma}\gamma^{\mu} &=0,\\
\gamma_{\mu}\gamma_{5}\gamma_{\rho}\gamma^{\mu} &=-2\gamma_{\rho}\gamma_{5}.\notag
\end{align}
We take two spinor $\psi$ and $\chi$ whose components anticommute. The Fierz identity
\begin{equation}
4\psi\bar\chi=-(\bar\chi\psi)-\gamma_{\mu}(\bar\chi\gamma^{\mu}\psi)
+\frac12\gamma_{\mu\nu}(\bar\chi\gamma^{\mu\nu}\psi)
+\gamma_{5}\gamma_{\mu}(\bar\chi\gamma_{5}\gamma^{\mu}\psi)
-\gamma_{5}(\bar\chi\gamma_{5}\psi)
\end{equation}
allows us to write the matrix $\psi\bar\chi$ as a linear combination of the antisymmetrized
products of $\gamma$ matrices. The following identities follow from the identity (A.5):
\begin{align}
\bar\psi\gamma^{\rho}\gamma_{\mu\nu}\chi&=\bar\chi\gamma_{\mu\nu}\gamma^{\rho}\psi,\\
\chi\bar\psi-\psi\bar\chi&=-\frac12\gamma_{\mu}(\bar\psi\gamma^{\mu}\chi)
+\gamma_{\mu\nu}(\bar\psi\gamma^{\mu\nu}\chi).
\end{align}
We define the hermitian conjugate as if the spinor components are operators in a Hilbert
space. For Majorana spinors we have
\begin{align}
({\bar\psi\gamma_{\mu_{1}\dots\mu_{n}}\chi})^{\dagger}
&=(-1)^{n}\bar\psi\gamma_{\mu_{1}\dots\mu_{n}}\chi,\\
\bar\psi\gamma_{\mu_{1}\dots\mu_{n}}\chi
&=(-1)^{n(n-1)/2}\,\bar\chi\gamma_{\mu_{1}\dots\mu_{n}}\psi.
\end{align}
In particular, $\bar\psi\gamma_{\mu}\psi=\bar\psi\gamma_{\mu\nu}\psi=0$. Setting $\chi=\psi$
in (A.5) we obtain for Majorana spinors the identity
\begin{equation}
\psi(\bar\psi\psi)+\gamma_5\psi(\bar\psi\gamma_5\psi)=0.
\end{equation}
In the paper we use a Majorana representation of the Dirac algebra in which the gamma matrices
are all imaginary and the spinors are real.

\end{document}